\documentclass[twocolumn]{aastex631}
\usepackage{graphicx,float}
\usepackage{scrextend}

\hypersetup{linkcolor=red,citecolor=cyan,filecolor=green,urlcolor=cyan}

\newcommand{\totarea}{1.43}
\newcommand{\dasharea}{1.35}
\newcommand{\cdasharea}{0.49}
\newcommand{\ancarea}{0.21}

\newcommand{\totpointings}{1912}

\newcommand{\dashpointings}{1256}
\newcommand{\totorbit}{157}
\newcommand{\cdashorbit}{57}

\newcommand{\psdepthcomb}{24.74$\pm$0.20}  
\newcommand{\psdepthdash}{24.72$\pm$0.17}  
\newcommand{\psdeptharchiv}{26.88$\pm$0.40}  

\newcommand{\npsfstar}{12335}  

\shorttitle{3D-DASH Imaging Survey}
\shortauthors{Mowla, Cutler, Brammer et al.}
\graphicspath{{./}{}}

\begin{document}

\title{3D-DASH: The Widest Near-Infrared Hubble Space Telescope Survey}

\author[0000-0002-8530-9765]{Lamiya A. Mowla}
\email{lamiya.mowla@utoronto.ca}
\affiliation{Dunlap Institute for Astronomy and Astrophysics, University of Toronto, 50 St George St, Toronto, ON M5S 3H4, Canada.}

\author[0000-0000-0000-0000]{Sam E. Cutler}
\affiliation{Department of Astronomy, University of Massachusetts, Amherst, MA 01003, USA}

\author[0000-0003-2680-005X]{Gabriel B. Brammer}
\affiliation{Cosmic Dawn Center (DAWN)}
\affiliation{Niels Bohr Institute, University of Copenhagen, Lyngbyvej 2, 2100 Copenhagen, Denmark}

\author[0000-0003-1665-2073]{Ivelina G. Momcheva}
\affiliation{Space Telescope Science Institute, Baltimore, MD 21218, USA}

\author[0000-0001-7160-3632]{Katherine E. Whitaker}
\affiliation{Department of Astronomy, University of Massachusetts, Amherst, MA 01003, USA}
\affiliation{Cosmic Dawn Center (DAWN)}

\author[0000-0002-8282-9888]{Pieter G. van Dokkum}
\affiliation{Astronomy Department, Yale University, New Haven, CT 06511, USA}

\author[0000-0001-5063-8254]{Rachel S. Bezanson}
\affiliation{Department of Physics and Astronomy and PITT PACC, University of Pittsburgh, Pittsburgh, PA 15260, USA}

\author[0000-0003-4264-3381]{Natascha M. F$\ddot{o}$rster Schreiber}
\affiliation{Max-Planck-Institut f{\"u}r extraterrestrische Physik (MPE), Giessenbachstr. 1, D-85748 Garching, Germany}

\author[0000-0002-8871-3026]{Marijn Franx}
\affiliation{Leiden Observatory, P.O. Box 9513, 2300 RA, Leiden, The Netherlands}

\author[0000-0001-9298-3523]{Kartheik G. Iyer}
\affiliation{Dunlap Institute for Astronomy and Astrophysics, University of Toronto, 50 St George St, Toronto, ON M5S 3H4, Canada.}

\author[0000-0001-9002-3502]{Danilo Marchesini}
\affiliation{Department of Physics and Astronomy, Tufts University, Medford, MA}

\author{Adam Muzzin}
\affiliation{Department of Physics and Astronomy, York University, 4700 Keele Street, Toronto, Ontario, ON MJ3 1P3, Canada}

\author{Erica J. Nelson}
\affiliation{Department for Astrophysical and Planetary Science, University of Colorado, Boulder, CO 80309, USA}

\author[0000-0001-7393-3336]{Rosalind E. Skelton}
\affiliation{South African Astronomical Observatory, P.O. Box 9, Observatory 7935, South Africa}

\author[0000-0002-4226-304X]{Gregory F. Snyder}
\affiliation{Space Telescope Science Institute, 3700 San Martin Dr, Baltimore, MD, USA 21218}

\author[0000-0000-0000-0000]{David A. Wake}
\affiliation{Department of Physics and Astronomy, University of North Carolina Asheville, Asheville, NC 28804, USA}

\author[ 0000-0003-3735-1931]{Stijn Wuyts}
\affiliation{Department of Physics, University of Bath, Claverton Down, Bath BA2 7AY, UK}

\author{Arjen van der Wel}
\affiliation{{Sterrenkundig Observatorium, Universiteit Gent, Krijgslaan 281 S9, B-9000 Gent, Belgium}}




\begin{abstract}
The 3D-Drift And SHift (3D-DASH) program is a \textit{Hubble Space Telescope} WFC3 F160W imaging and G141 grism survey of the equatorial COSMOS field. 
3D-DASH extends the legacy of HST near-infrared imaging and spectroscopy to degree-scale swaths of the sky, enabling the identification and study of distant galaxies ($z>2$) that are rare or in short-lived phases of galaxy evolution at rest-frame optical wavelengths. Furthermore, when combined with existing ACS/F814W imaging, the program facilitates spatially-resolved studies of the stellar populations and dust content of intermediate-redshift ($0.5<z<2$) galaxies.
Here we present the reduced F160W imaging mosaic available to the community. Observed with the efficient DASH technique, the mosaic comprises \dashpointings\ individual WFC3 pointings, corresponding to an area of \dasharea\ deg$^2$ (\totarea\ deg$^2$ in \totpointings\ when including archival data). The median $5\sigma$ point-source limit in $H_{160}$ is \psdepthcomb\ mag. We also provide tools to determine the local point spread function (PSF), create cutouts, and explore the image at any location within the 3D-DASH footprint\footnote{\url{https://archive.stsci.edu/hlsp/3d-dash/}}\footnote{\url{www.lamiyamowla.com/3d-dash}}. 3D-DASH is the widest \textit{HST}/WFC3 imaging survey in the F160W filter to date, increasing the existing extragalactic survey area in the near-infrared at HST resolution by an order of magnitude.
\end{abstract} 

\accepted{for publication by Astrophysical Journal}


\vspace{0.15cm}  

\section{Introduction} 
\label{sec:intro}

Wide-field near-infrared (NIR) surveys have proven invaluable for the study of the high mass end of the galaxy stellar mass function at $z>1$ (where rest-frame optical emission shifts into the NIR) and for determining the prevalence of short-lived events such as mergers, the properties and demographics of AGN, and the evolution of galaxy groups and clusters. Such surveys have been undertaken from the ground (e.g., NMBS \citet{Whitaker2011}, UltraVISTA \citet{McCracken2012,Muzzin2013}, UKIDSS-UDS \citet{Lawrence2007, Williams2009}), but so far not with the \textit{Hubble Space Telescope} (HST). 

Until recently, the largest area imaged with HST in the NIR ($J_{125}$ and $H_{160}$) was the 0.2 deg$^2$ CANDELS survey \citep[900 orbits;][]{Grogin2011,Koekemoer2011}. 
``Classical'' HST NIR surveys over larger areas have not been executed owing to the fact that a guide star acquisition is required for each new Wide Field Camera 3 (WFC3) pointing. In practice, this sets the minimum exposure time to 1 orbit per pointing, and this means that very large allocations are needed to cover substantial ($>0.5$ deg$^2$) areas. This situation changed with the development and implementation of the Drift And SHift technique \citep{Momcheva2017}, which enables up to 8 pointings to be observed within a single orbit (see \S\,2.1). This technique allows for an order of magnitude increase in the efficiency of large-area mapping with HST.

The obvious field for a wide-area HST survey in the NIR is
the ACS COSMOS field \citep[640 orbits; ][]{Scoville2007,Koekemoer2007} covering 1.7 degrees$^2$. Fifteen years after its execution this is still the most comprehensive and widest tier of the ``wedding cake" of extragalactic survey fields. COSMOS is the ultimate field to study rare objects such as the most massive galaxies \citep{Hill2017,mowladash2019,Marsan2019,Cooke2019}, close pairs \citep{Xu2012}, overdense structures \citep{Chiang2014,Iovino2016}, active galactic nuclei (AGN) \citep{Civano2012}, post-starburst galaxies \citep{WhitakerPSB2012}, etc. The field has exquisite ancillary data ranging from X-rays \citep[e.g.][]{Civano2016,Marchesi2016}, to submillimeter \citep[e.g.][]{Faisst2020,LeFevre2020}, radio \citep[e.g.][]{Fernandez2016}, and in spectroscopy \citep[e.g.][]{vanderwel2016}. It has provided the best constraints that we currently have on the evolution of the galaxy stellar mass function \citep{Muzzin2013}, the evolution of the star formation rates of galaxies \citep{Scoville2013,Lee2015}, and the evolution of their sizes \citep{Sargent2007}. However, all these measurements suffer from biases as the field lacks high-resolution near-infrared imaging. 

In HST Cycle 23 we obtained pilot DASH observations in a section of the COSMOS field, using \cdashorbit\ orbits to cover \cdasharea\ deg$^2$. Several scientific results validated the success of the DASH technique. The COSMOS-DASH image extended the parameter space probed for the galaxy size-mass relation and found that there is no significant difference between the sizes of star-forming and quiescent galaxies at the high mass end \citep{mowladash2019}, the size-mass relation of all galaxies can be described by a broken power-law with a pivot that is reminiscent of the stellar mass-halo mass relation of galaxies \citep{mowla2019smhm}, and that there is a flattening of the quiescent size-mass relation at lower stellar masses \citep{Nedkova2021,Cutler2022}. COSMOS-DASH has also been used to find early universe massive galaxy candidates \citep{Marsan2022}, which will be of particular importance during the \textit{James Webb Space Telescope} era.   

\begin{figure*}[t]
\centering
\includegraphics[width=0.47\textwidth]{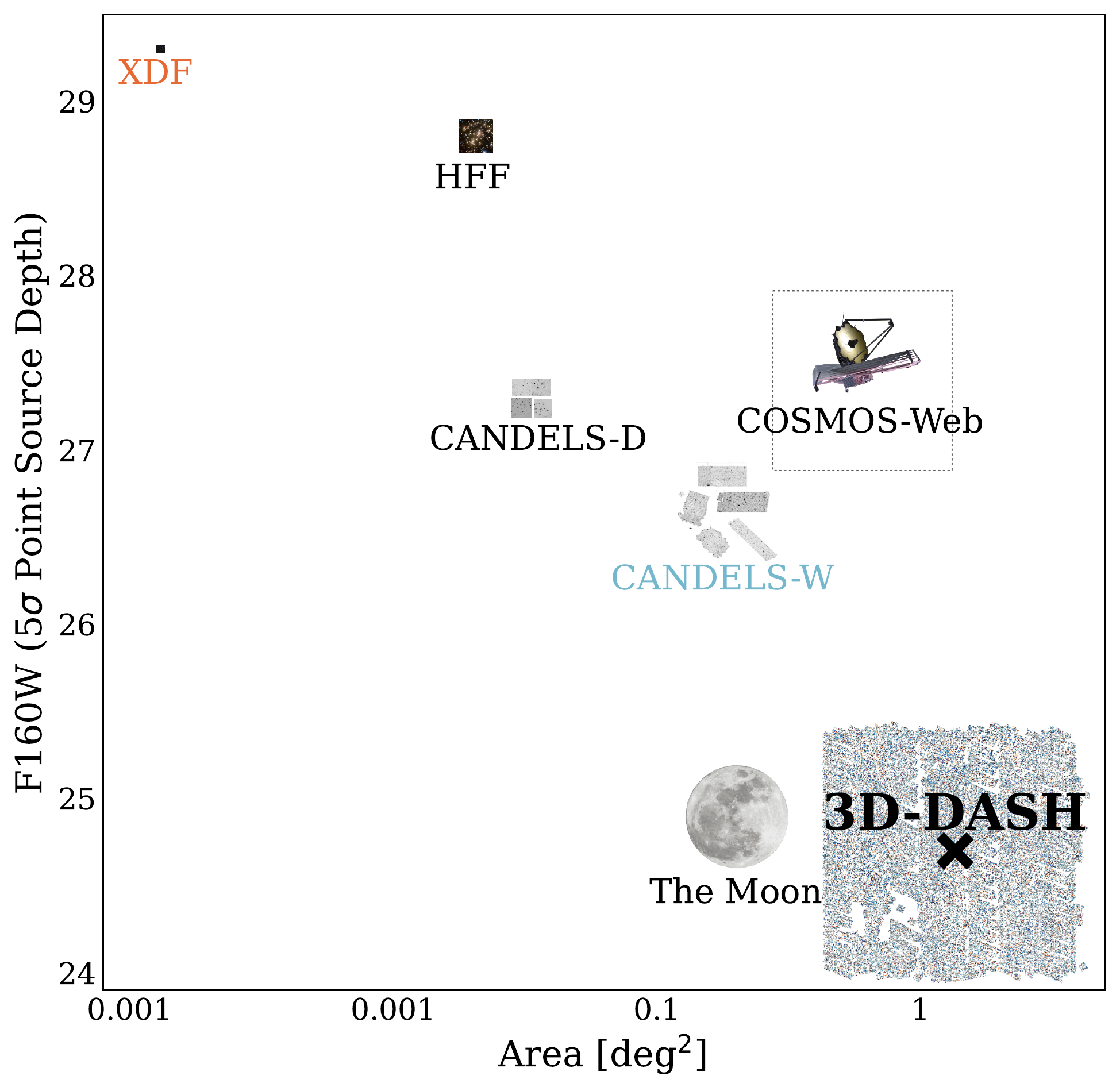}
\includegraphics[width=0.49\textwidth]{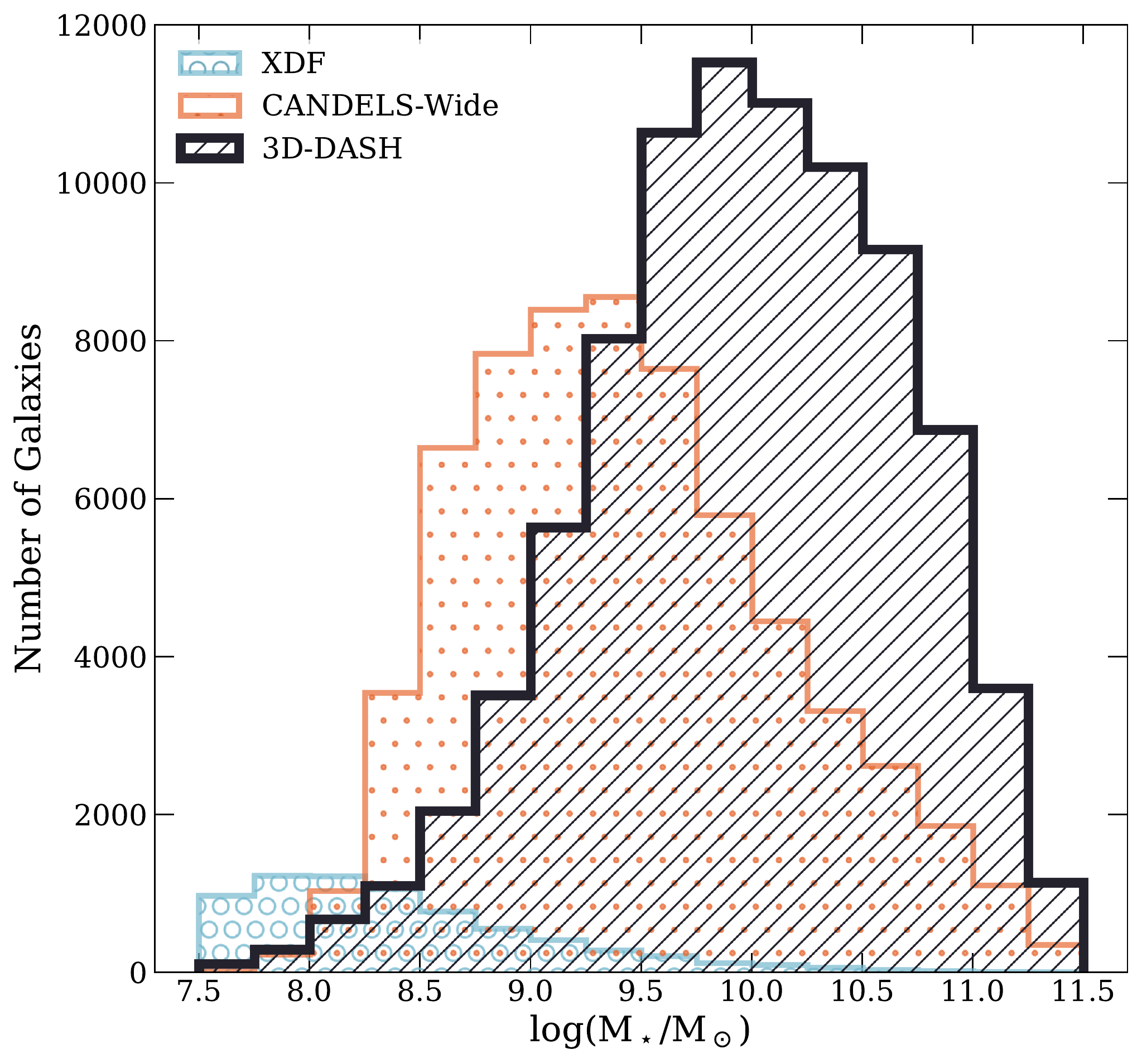}
\caption{Left: Area vs. depth for WFC3/F160W imaging surveys. The area of the thumbnails is proportional to the area of the surveys, which includes the Hubble Extreme Deep Field \citep[XDF, 330 orbits, ][]{xdf2013}, Hubble Frontier Fields \citep[][]{lotz2017,shipley2018}, and the Cosmic Assembly Near-Infrared Deep Extragalactic Legacy Survey \citep[CANDELS, 900 orbits ][]{Koekemoer2007}. The upcoming COSMOS-Web survey \citep[][]{cosmoswebb2021} with the \textit{James Webb Space Telescope} \citep[JWST][]{jwst2006} is shown by the open box with JWST. 
The 3D-DASH survey, observed using the Drift And SHift technique \citep{Momcheva2017}, is shown with a black cross. It subsumes COSMOS-DASH \citep[\cdashorbit\ orbits observed with DASH, ][]{mowladash2019} and other ancillary data in the COSMOS field, which are listed in Table \ref{tab:archival}. Right: The number of galaxies per stellar mass bins that are observed within three tiers of the HST extragalactic wedding-cake: XDF (Iyer, Mowla in prep., \citet{Sorba2018}), 3D-HST/CANDELS \citep[][]{Leja2020} and 3D-DASH \citep[COSMOS2020][]{Weaver2022}. For XDF and 3D-HST/CANDELS, we have selected all galaxies from their respective catalogs within $0.1<z<3$. For 3D-DASH we have selected galaxies from COSMOS2020 catalog with $H<23$ within $0.1<z<3$ (requirement for size measurement with $\delta r_e/r_e<0.2$ from \citet{Cutler2022}). We note that these different surveys are not calibrated to match, except for the use of the same IMF, and thus we expect systematic uncertainties in stellar mass on the order of 0.2 dex between the samples that are not corrected here.}
\label{fig:dash_depth_area}
\end{figure*}

\begin{figure*}[t]
\centering
\includegraphics[width=0.98\textwidth]{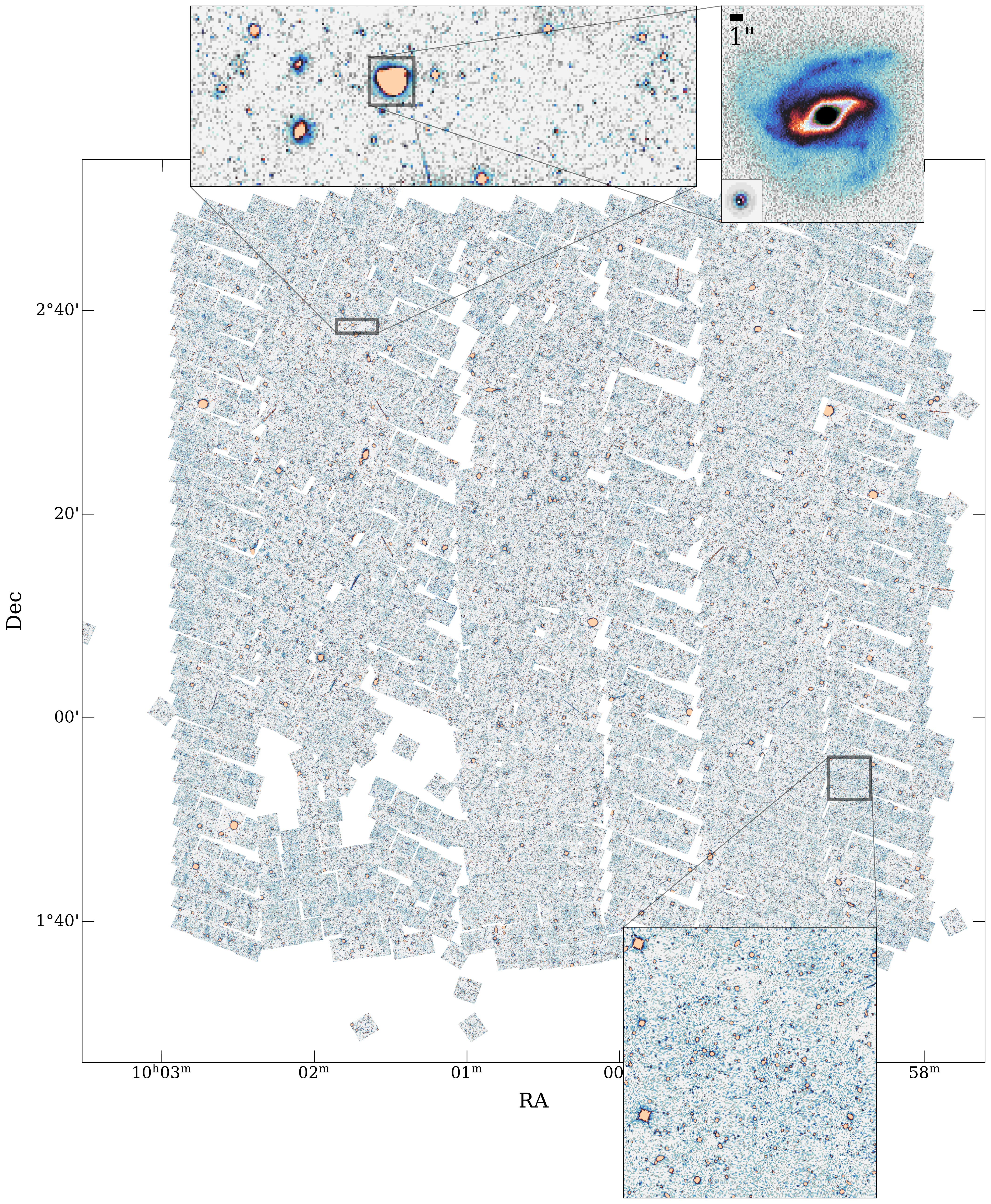}
\caption{The 3D-DASH F160W mosaic. The area contributed by 3D-DASH is \dasharea\,deg$^2$; the total area with all ancillary F160W data in the COSMOS field, including the deeper CANDELS imaging and various other archival data sets listed in Table \ref{tab:archival}, is \totarea\,deg$^2$. The zoomed-in panels reveal the wealth of bright objects that can be studied in this high resolution shallow tier of the extragalactic wedding cake. The point spread function at the location of the zoomed-in galaxy is shown in the inset.}
\label{fig:dash_img}
\end{figure*}

\begin{figure*}[t]
\centering
\includegraphics[width=\textwidth]{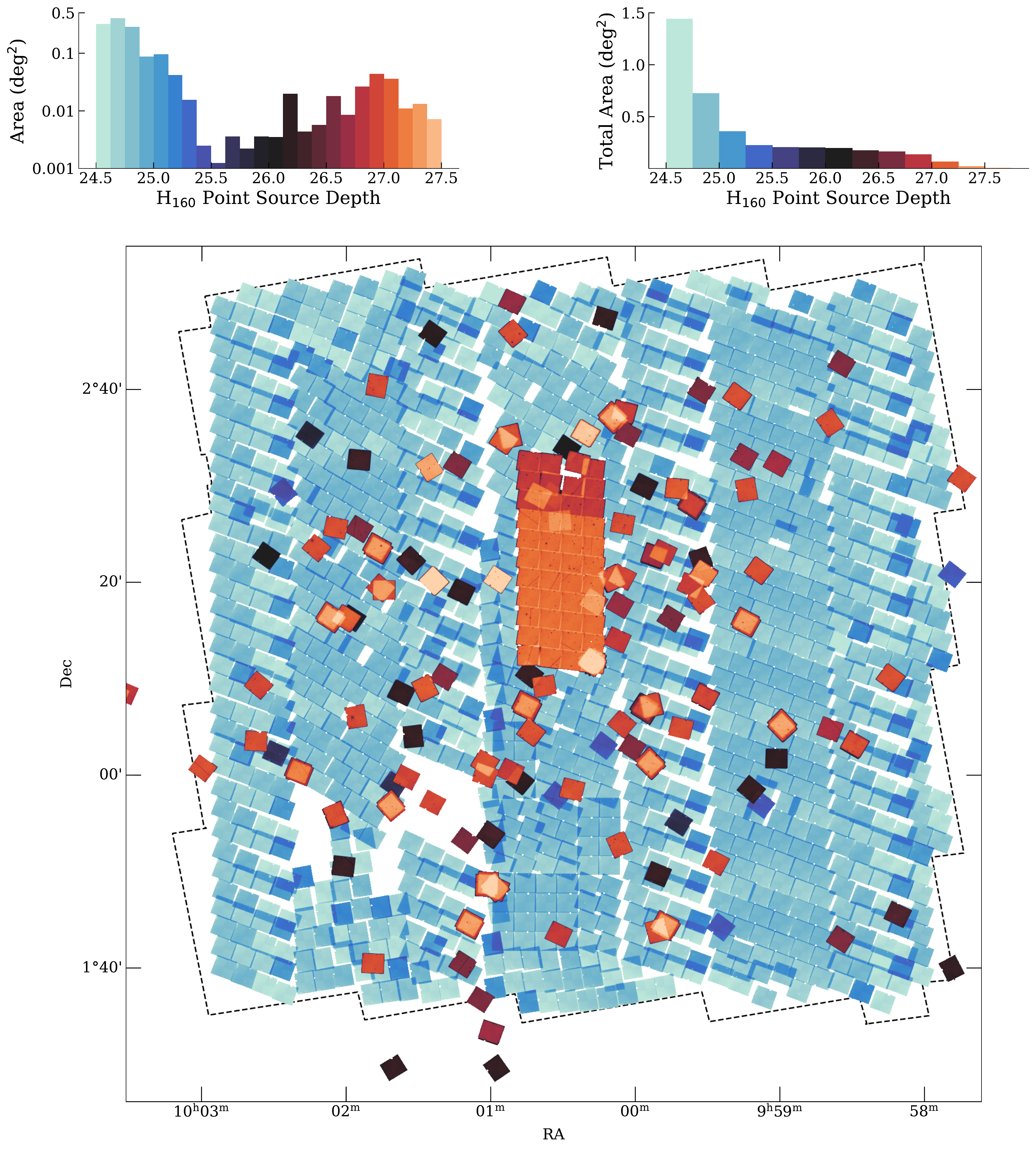}
\caption{Bottom: The 3D-DASH $H_{160}$ point source depth map. The area contributed by 3D-DASH is \dasharea\,deg$^2$ shown by the blue pointings. The total area is \totarea\,deg$^2$ and includes all ancillary data in the COSMOS field shown by black to orange pointings, including the deeper CANDELS imaging and various other archival data sets listed in Table \ref{tab:archival}. The dashed-black outline shows the 1.64 deg$^2$ covered by ACS/F814W \citep{Koekemoer2007}. Top left: Distribution of point source depths in 3D-DASH mosaic, with the wider and shallower images coming from the DASH-programs in Cycle 23 and 28, and the deeper images coming from CANDELS and other ancillary programs. Top right: The cumulative area available in 3D-DASH below a certain point certain depth.}
\label{fig:dash_wht}
\end{figure*}

3D-DASH increases coverage to the entire ACS COSMOS field with both imaging  and spectroscopy, thus extending the legacy of HST into the regime of degree-scale NIR imaging and spectroscopy. Combining the pilot program with new data taken in Cycles 23 and 28, the ACS COSMOS field now has coverage in both WFC3/F160W and WFC3/G141, widening the shallow base of the extragalactic wedding cake (Figure \ref{fig:dash_depth_area} and Table \ref{tab:summary}). 3D-DASH covers \dasharea\ deg$^2$ of the COSMOS field down to median depth of $H_{160}=$ \psdepthdash\, or \totarea\ deg$^2$ down to $H_{160}=$ \psdepthcomb\ when ancillary data are included, increasing the extragalactic survey area observed by HST in the NIR by an order of magnitude\footnote{All our data products are available at MAST as a High Level Science Product via \dataset[10.17909/srcz-2b67]{\doi{10.17909/srcz-2b67}} and here \url{https://archive.stsci.edu/hlsp/3d-dash/}. We also provide tools to generate local point spread functions and make cutouts at any location within the 3D-DASH footprint.}. This paper presents the F160W imaging of this program (Figures \ref{fig:dash_img} and \ref{fig:dash_wht}), while a forthcoming paper will present the grism spectroscopy (Figure \ref{fig:grism}). 


The paper is structured as follows: in Section \ref{sec:dash} we give a description of the 3D-DASH survey, Section \ref{sec:image} describes the image processing, Section \ref{sec:example} highlights the potential of the data with three examples of science applications, and Section \ref{sec:summary} gives a summary of the paper. We assume a $\Lambda$CDM cosmology with $\Omega_{\rm m}=$ 0.3, $\Omega_{\Lambda}=$ 0.7, and $H_0=$ 70 km s$^{-1}$ Mpc$^{-1}$. The AB magnitude system \citep{Oke1983} is adopted throughout the paper; a Chabrier (2003) initial mass function (IMF) is used when appropriate.

\begin{table*}[t]
\centering
\begin{tabular}{p{0.4\linewidth}  p{0.35\linewidth}}
\multicolumn{2}{c}{3D-DASH IMAGING SUMMARY}                        \\ \hline
\hline
Position (J2000)      & R.A. 10h 00m 25.4s, Dec. +02$^\circ$ 12$^\prime$ 21.\farcs0 \\
DASH-only area & \dasharea\ deg$^2$                     \\
Ancillary area & \ancarea\ deg$^2$  \\
Area (combined)          & \totarea\   deg$^2$             \\
Instrument/Filter     & WFC3/F160W                       \\
Exposure Dates (DASH)        &  November 2016-June 2017 (Cycle 23) \\ 
& December 2020-March2022 (Cycle 28)        \\
Exposure Dates (Ancillary)        &  2010 - 2022                                  \\
Number of DASH pointings   &  \dashpointings\                                  \\
Zero point $H_{160}$ & 25.95 \\
Pixel scale & 0\farcs1 \\
Median DASH-only depth\footnote{\label{foottab1}5 $\sigma$ $H_{160}$ point source depth}       &  \psdepthdash\           \\
Median ancillary depth\footref{foottab1}       &  \psdeptharchiv\           \\
Median 3D-DASH depth\footref{foottab1} (including ancillary)     &   \psdepthcomb\                            \\
DOI & \dataset[10.17909/srcz-2b67]{\doi{10.17909/srcz-2b67}}  \\
Archive Link          &  \url{https://archive.stsci.edu/hlsp/3d-dash}                               \\ \hline
\end{tabular}
\caption{Summary of the imaging data released in the 3D-DASH$\_v1.0$.  }
\label{tab:summary}
\end{table*}

\vspace{0.5cm}
\section{The 3D-DASH Program}
\label{sec:dash}


\subsection{Drift And SHift (DASH)}





Drift And SHift (DASH) is an efficient technique to cover extensive area using the near-infrared (IR) channel of Wide Field Camera 3 (WFC3) on the \textit{Hubble Space Telescope} \citep{Momcheva2017}. In standard HST observations, guide stars are acquired for each new pointing. Acquiring a guide star takes approximately 10 minutes, which means that short exposures are only possible with huge overheads. This limits the total number of pointings that can be obtained within an orbit's visibility window\footnote{HST is limited to two acquisitions per orbit by policy, see  \href{https://hst-docs.stsci.edu/stisihb/chapter-9-overheads-and-orbit-time-determination/9-2-stis-exposure-overheads}{STIS Instrument Handbook}}. 

The guide star acquisition limitation can be circumvented by acquiring a guide star for only the first pointing of an orbit and guiding with the three HST gyros for the remainder. This technique makes it possible to observe up to 8 WFC3 pointings in a single orbit, significantly increasing HST's large-scale mapping capabilities. Successfully used for the first time in the COSMOS-DASH survey \citep{mowladash2019} to image 0.49 deg$^2$ of the three stripes of the UltraVISTA field \citep{McCracken2012}, DASH remains the most efficient method for wide-area observations with WFC3. 

During a standard guided exposure, the three HST gyros receive continuous corrections from the Fine Guidance Sensors (FGS). Turning off guiding stops the stream of corrections from the FGS, and the telescope begins to drift with an expected rate of $0\farcs 001 - 0\farcs 002$ per second. In CCDs, this would lead to detrimental smearing of the image in a typical 5-minute exposure. However,
the WFC3/IR detector can perform multiple non-destructive, zero overhead reads throughout the exposure. Setting the time between reads to 25 seconds or less lowers the drift between reads to $\leq 0\farcs12$ - around a pixel (pixel scale $=0\farcs129$). The data obtained between the reads can be treated as independent 25\,s exposures that can be drizzled to restore the full resolution of WFC3. In \citet{Momcheva2017} and \citet{mowladash2019} we demonstrated that the resolution of the WFC3 camera is preserved in this process, and the measured structural parameters of the galaxies are consistent with those measured in guided observations \citep{Cutler2022}.

\subsection{Observations}

The DASH method was used in the Cycle 23 and 28, 3D-DASH program (Program ID: GO-14114 and GO-16259) to obtain \dashpointings\ WFC3 $H_{160}$ pointings in \totorbit\ orbits, covering an area of \dasharea\ deg$^2$ in the COSMOS field. The design of the mosaic was constrained by the available orientations. To ease scheduling no initial ORIENT constraints were imposed. After scheduling windows were assigned the final mosaic was designed, constrained by the ORIENTs that were allowed within each of the windows. A consequence of this strategy is that the final mosaic has gaps, as the wide range of orientations
precluded the design of a fully contiguous mosaic that also avoids severe overlaps between pointings.
The Cycle 23 data were obtained between November 2016 and June 2017, while the Cycle 28 data were obtained between December 2020 and March 2022. 

Each orbit consists of a guide star acquisition, followed by a single guided exposure. Next, the stream of corrections from the fine guidance sensor is turned off, and the telescope is moved to the second exposure. Guiding remains off for the next six exposures. The exposure times, and hence the number of independent reads within each exposure, were determined by the requirement to fit eight pointings in a single orbit. 

Following the 2018 failure of one of the HST gyros, a new, previously dormant, gyro was turned on. The new gyro exhibited extremely high bias at the start of operations, which necessitated certain changes in the observatory controls which led to differences between the GO-14114 (executed prior to the gyro switch) and GO-16259 (executed post-switch) observations. The orbital visibility in COSMOS  post-switch is shorter by 290 seconds, the duration of guide star acquisitions and commanded offsets is increased and the observatory is not allowed to execute an offset during the five minutes after the guide star acquisition. The combined effect is a shorter length of the science exposures. With the additional constraints of the SPARS25 sampling sequence (exposure times must change by integer factors of 25 seconds) and avoiding buffer dump interruptions, we find a new optimum sequence of exposure times. The total exposure time in the orbit is 350 seconds shorter than in GO-14114 (1774 seconds vs. 2124 seconds). Figure \ref{fig:struct} and Table \ref{tab:struct} show the APT graphical representation of a single GO-16259 orbit and the sequence of events, respectively. These can be compared to Figure 2 and Table 1 in \citet{Momcheva2017}. 

\begin{figure}[t]
\centering
\includegraphics[width=0.48\textwidth]{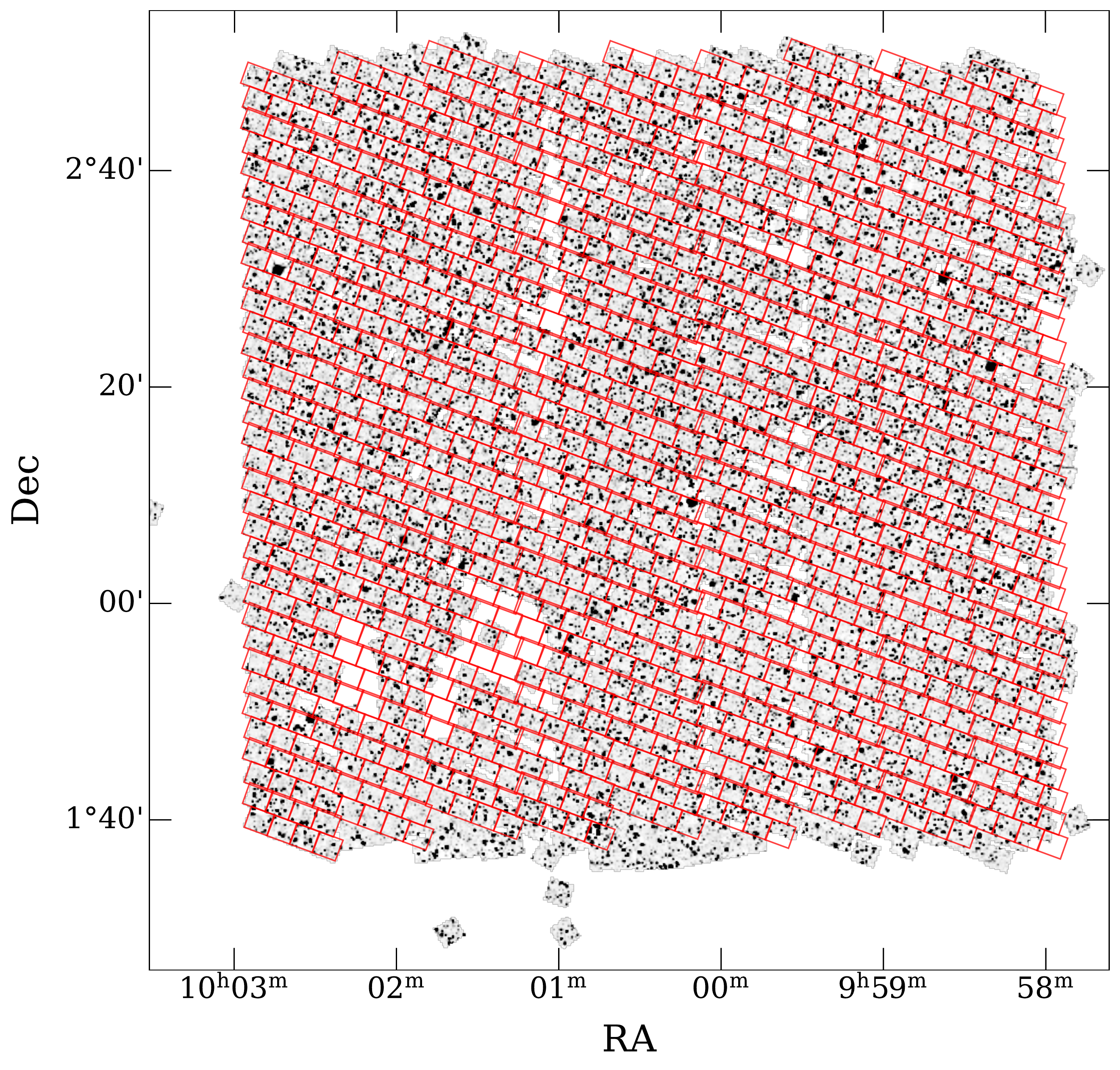}
\caption{The 3D-DASH \textit{WFC3/G141} grism spectroscopy footprint covering the COSMOS field. In Cycle 28  and 29, grism observations of \totarea\ deg$^2$ area will be obtained in 1272 pointings in 157 HST orbits using the Drift And SHift technique.}
\label{fig:grism}
\end{figure} 

The 3D-DASH WFC3/F160W imaging will be supplemented by WFC3/G141 grism spectroscopy, executed in Cycle 28 and 29 in 159 orbits (see Figure \ref{fig:grism} for the grism observation footprint). 3D-DASH will increase the area observed with infrared grism spectroscopy by almost an order of magnitude and map the spatially resolved emission lines of over thousand star-forming galaxies at cosmic noon, providing crucial information on the build up of disks and bulges of galaxies. Similar to the imaging, the grism spectroscopy will cover the the ACS-COSMOS field, observed efficiently in 1272 pointings using the DASH technique. 


\section{The 3D-DASH Imaging Mosaic}
\label{sec:image}

\subsection{Preparation of 3D-DASH mosaic}
\label{ap:reduction}

Owing to the shifts during the exposures, the reduction of DASH data is more complex than that of guided exposures. Here we provide a summary of the reduction procedures; we also refer to \citet{Momcheva2017} where the reduction of a subset of the 3D-DASH\footnote{We refer to the full combined dataset as 3D-DASH, subsuming the pilot COSMOS-DASH data set.} data were first described.

Each DASH orbit consists of one guided and seven unguided exposures with offsets of roughly 2$\arcmin$ between each. \citet{Momcheva2017} describe the overall mosaic strategy and detailed commanding instructions for the DASH visits.  For uniformity of the data reduction, we process both guided and unguided exposures using the same analysis pipeline as outlined below.  

The raw WFC3/IR images were downloaded from the Mikulski Archive for Space Telescopes (MAST\footnote{\url{http://archive.stsci.edu}}) and processed into calibrated exposure ramps (``IMA'' products) with the \texttt{calwf3} pipeline after disabling the pipeline cosmic ray identification step (\texttt{CRCORR=False}). The IMA files provide a measure of the total charge on the detector sampled every 25 seconds over the duration of the exposure (253 or 278 seconds for exposures with \texttt{NSAMP}=12 and 13, respectively).  To reduce the degradation of the image quality of a given exposure due to the telescope drifts, we take \textit{image differences} up the ramp and generate $\mathtt{NSAMP}-2$ essentially independent calibrated exposures with drifts now integrated over 25 seconds rather than the full exposure duration\footnote{Software to split DASH exposures into difference images is provided at https://github.com/gbrammer/wfc3dash}.  The properties of these difference images are essentially identical to normal calibrated WFC3/IR ``FLT'' products, though with slightly different noise characteristics.  Taking image differences increases the effective read noise by a factor of $\sqrt{2}$ and the read noise of two adjacent image differences will be anti-correlated as the measured (noisy) flux of a given read appears as negative in the first difference image and positive in the second.  In the case of guided exposures with no drifts, the image differences are equivalent to taking the pixel values of the last read minus the first, with read noise from just those two reads.  The differences do not cancel out for sequences with drifts, where the pixel indices of the difference images are effectively shifted when they are combined into the output mosaic.

The DASH observations produce $N\times 7 \times (\mathtt{NSAMP}-2)$ difference image ``exposures'', with $N$ the number of orbits.  These exposures are  processed in an identical way with the $N$ guided exposures that are taken at the start of each orbit, and with all guided archival observations in the COSMOS field.  We first compute an internal alignment of the visit exposures using sources detected in the images (both stars and galaxies), which corrects the DASH drifts between samples and small pointing errors typical of the guided sequences. We generate a small mosaic of the visit exposures to detect fainter sources, and align these mosaics to galaxies in the F814W catalogs provided by the COSMOS collaboration \citep{Koekemoer2007}.  Point sources are excluded from the catalog alignment as stars can have significant proper motions between the ACS-COSMOS and DASH epochs.  Since we do not identify cosmic rays in the DASH exposures at the pipeline level as with normal WFC3/IR exposures, we detect and mask the cosmic rays using the standard tools of the \texttt{AstroDrizzle} \citep{Gonzaga2012} package when creating the combined visit/pointing image (turning on cosmic ray identification is useful even for the guided exposures to mask unflagged hot pixels and weaker cosmic rays missed by \textit{calwf3}).  
In the pilot COSMOS-DASH program, several sequences of eight pointings were broken up due to South Atlantic Anomaly (SAA) passages. There was typically a large offset of $10\arcsec - 15\arcsec$ between the ``before" and ``after" exposures, due to the spacecraft drift during the SAA passage; nevertheless, no observable degradation of the PSF was found in these sequences. However, the interruption of DASH orbits by SAA passages lowered the efficiency of the observatory because the orbit following the SAA passage was only $1/4$ full. For this reason the planning team recommended that we avoid SAA passages and schedule all eight pointings within a single orbit. 

The requirement of three functioning HST gyros is crucial for optimum execution of the DASH technique. In recent years, multiple gyro failures have made DASH observations challenging. Here we show that the average drift between reads has increased from 0\farcs02 in the pilot COSMOS-DASH survey in 2017 to 0\farcs15 in 2021 and is over 0\farcs5 per read for $\sim$ 5\% of pointings as shown in Figure \ref{fig:drift}. The exact cause of this increase is not clear. This increased drift has resulted in a more irregular overlapping of pointings and some noticeable gaps in the mosaic, as can be seen in Figure \ref{fig:dash_wht}. It has also resulted in increased background noise and thus a lower point source depth than the pilot COSMOS-DASH program (see Section \ref{sec:psd}).


The full set of $\sim 5000$ aligned DASH exposures, together with all other existing $H_{160}$ data in the COSMOS field, are then drizzled into a single, large mosaic using \texttt{Astrodrizzle}. The weight map scales with both the variations in exposure time and the impact of drifting across the final mosaic, where lower weight corresponds to higher drift and/or shorter exposure times.  Both the science image and weight maps are drizzled to a pixel scale of $0\farcs 1$ using a square kernel and with \texttt{pixfrac}$=$0.8. The final mosaic of the 3D-DASH image is $53248\times 53248$ pixels and is centered at RA$=$10:00:28.6, DEC$=+$02:12:21.0.
The image is shown in Figure \ref{fig:dash_img} and the point source depth map created from the weight map (using the average noise in $0\farcs 3$ diameter apertures, details in Section \ref{sec:noise}) is shown in Figure \ref{fig:dash_wht}. All our data products are available at MAST as a High Level Science Product via \dataset[10.17909/srcz-2b67]{\doi{10.17909/srcz-2b67}} and here \url{https://archive.stsci.edu/hlsp/3d-dash/}. The science images their corresponding weight maps are available in three formats: 
\begin{enumerate}
    \item Full 3D-DASH mosaic\footnote{\label{3ddashimage} Combined DASH pointings and ancillary data.}  
    \item 3D-DASH mosaic\footref{3ddashimage} divided into 256 (16 $\times$ 16) tiles
    \item Full DASH-only mosaic 
\end{enumerate}

\begin{figure}[t]
\centering
\includegraphics[width=0.48\textwidth]{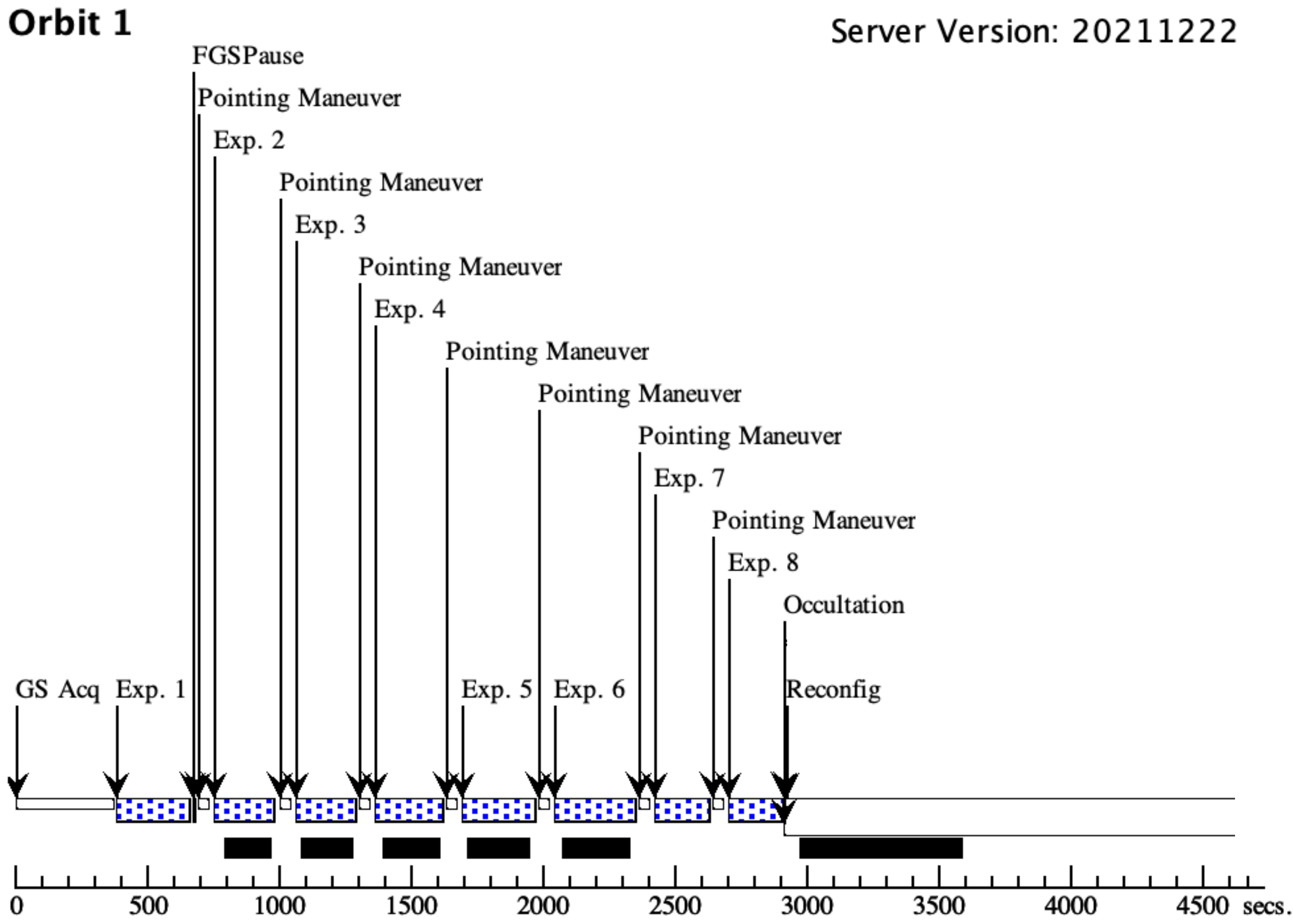}
\caption{Graphical representation of a single Drift And SHift orbit (Visit 1 of GO-16259). Only the first pointing is guided (PCS Mode=FINE). Blue, dotted bars indicate science exposures. Black bars are buffer dumps. The total exposure time is 1774 s, corresponding to 56\% of the total orbital visibility.}
\label{fig:struct}
\end{figure}

\begin{figure*}[htbp]
\centering
\includegraphics[width=0.45\textwidth]{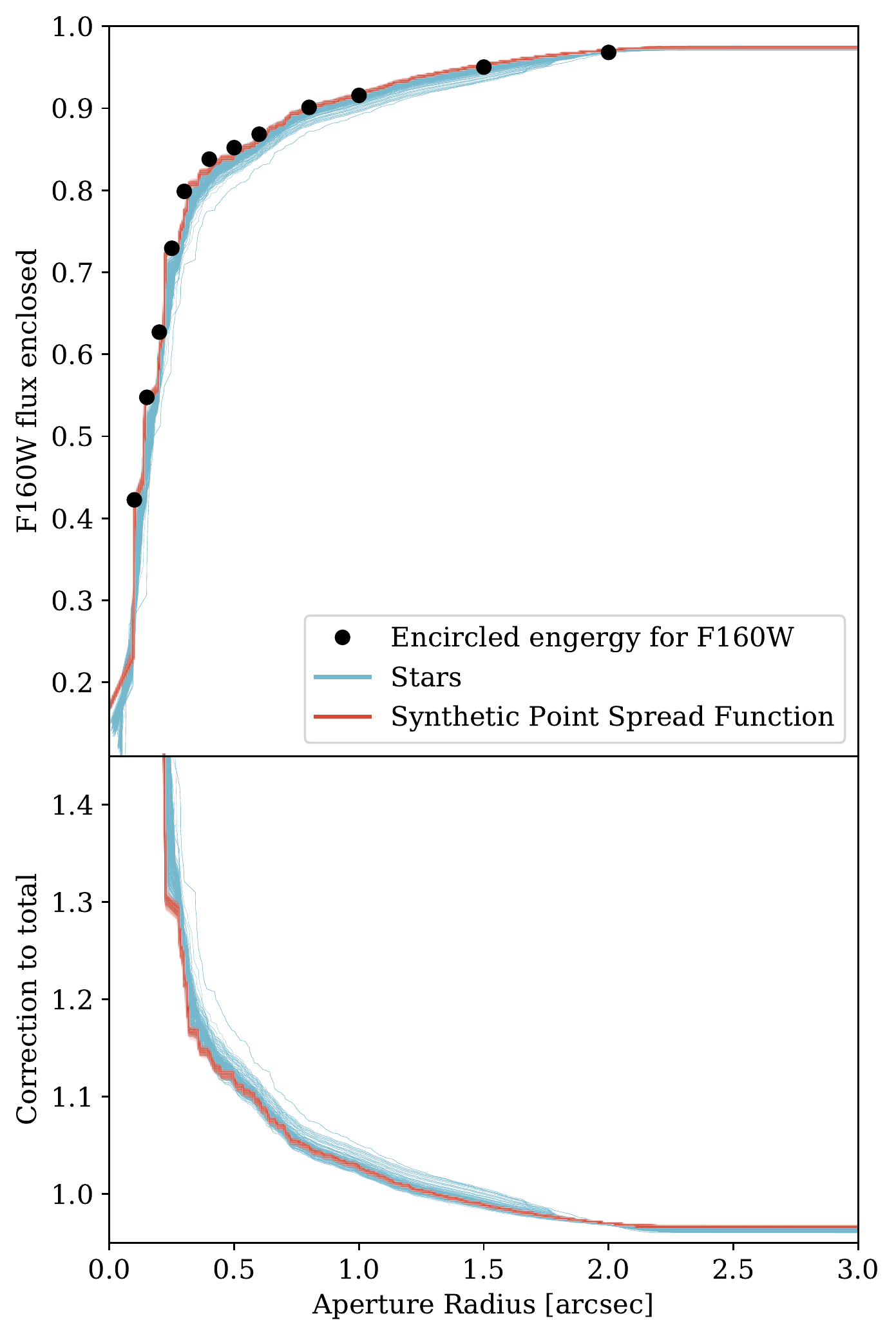}
\includegraphics[width=0.54\textwidth]{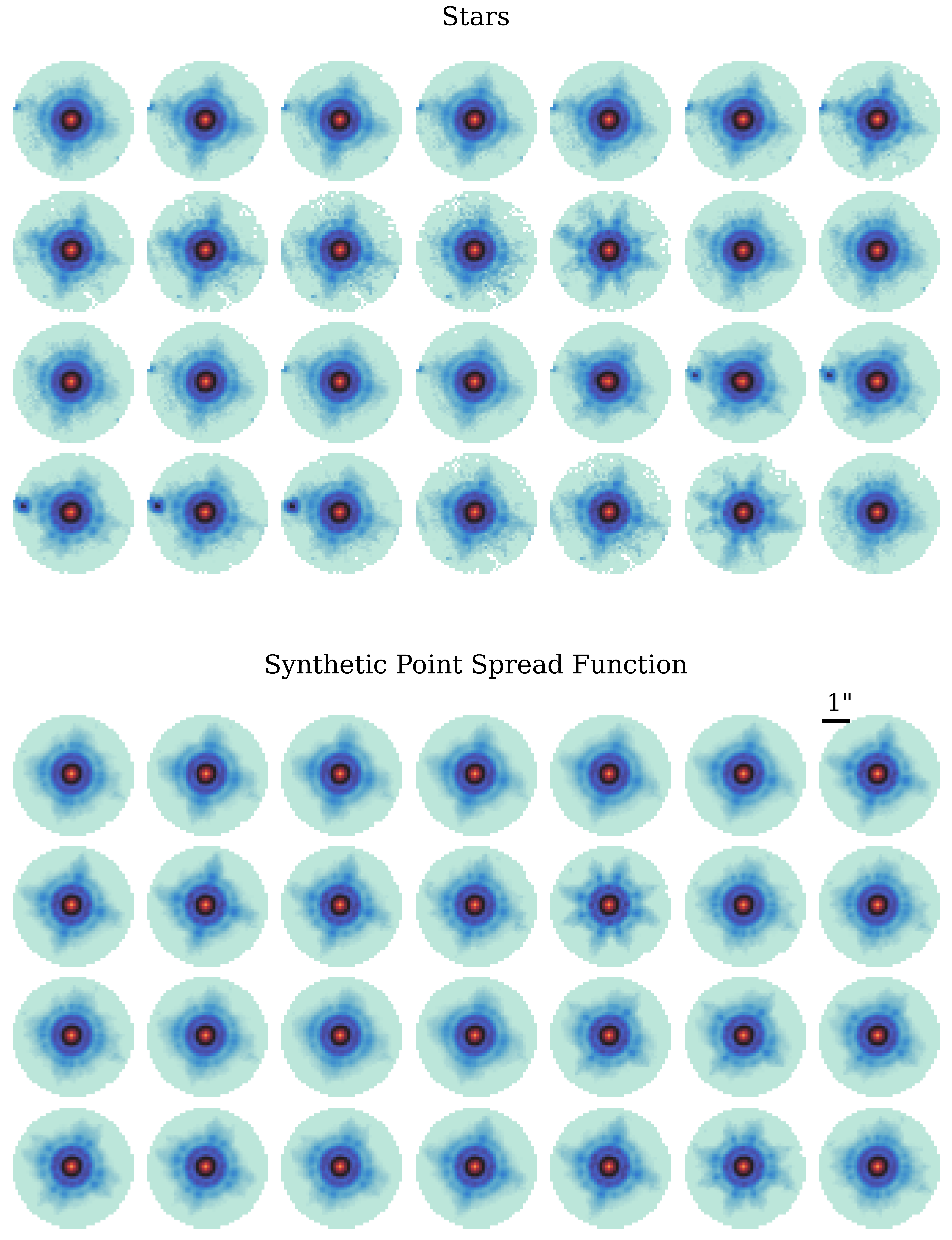}
\caption{The point-spread function (PSF) and growth-curve of 3D-DASH. Right: Example point spread functions from the 3D-DASH field. The top panel shows the stars selected from the Hyper Suprime-Cam Subaru Strategic Program \citep{Aihara2021} while the bottom panel shows synthetic position based point spread functions. Left: $H_{160}$ growth curve. Upper panel shows the fraction of total light enclosed as a function of radius from all the used $H_{160}$ PSF stamps. This is calculated as the fraction of light inside a given radius to the light inside 2$\arcsec$, $f(r)/f(2\arcsec)$, re-normalized to the fraction of total light enclosed within 2$\arcsec$ (97\% from WFC3 Instrument Handbook). (The blue lines show the growth curve of stars while the red lines are of the synthetic PSFs. The PSFs of the entire field are consistent with each other as shown by the blue and red lines. The black points show the encircled energy as a function of aperture size, also normalized to 2$\arcsec$, from the WFC3 Instrument Handbook. The lower panel shows the correction to total flux for point sources across the mosaic with a circularized Kron radius equal to the aperture radius on the x-axis. This is the inverse of the growth curves show in the upper panel ($f(2\arcsec)/f(r)$). The minimum Kron radius is set to 0.3$\farcs$, which requires a maximum correction of 1.4. }
\label{fig:psf}
\end{figure*}

\begin{figure*}[ht]
\centering
\includegraphics[width=\textwidth]{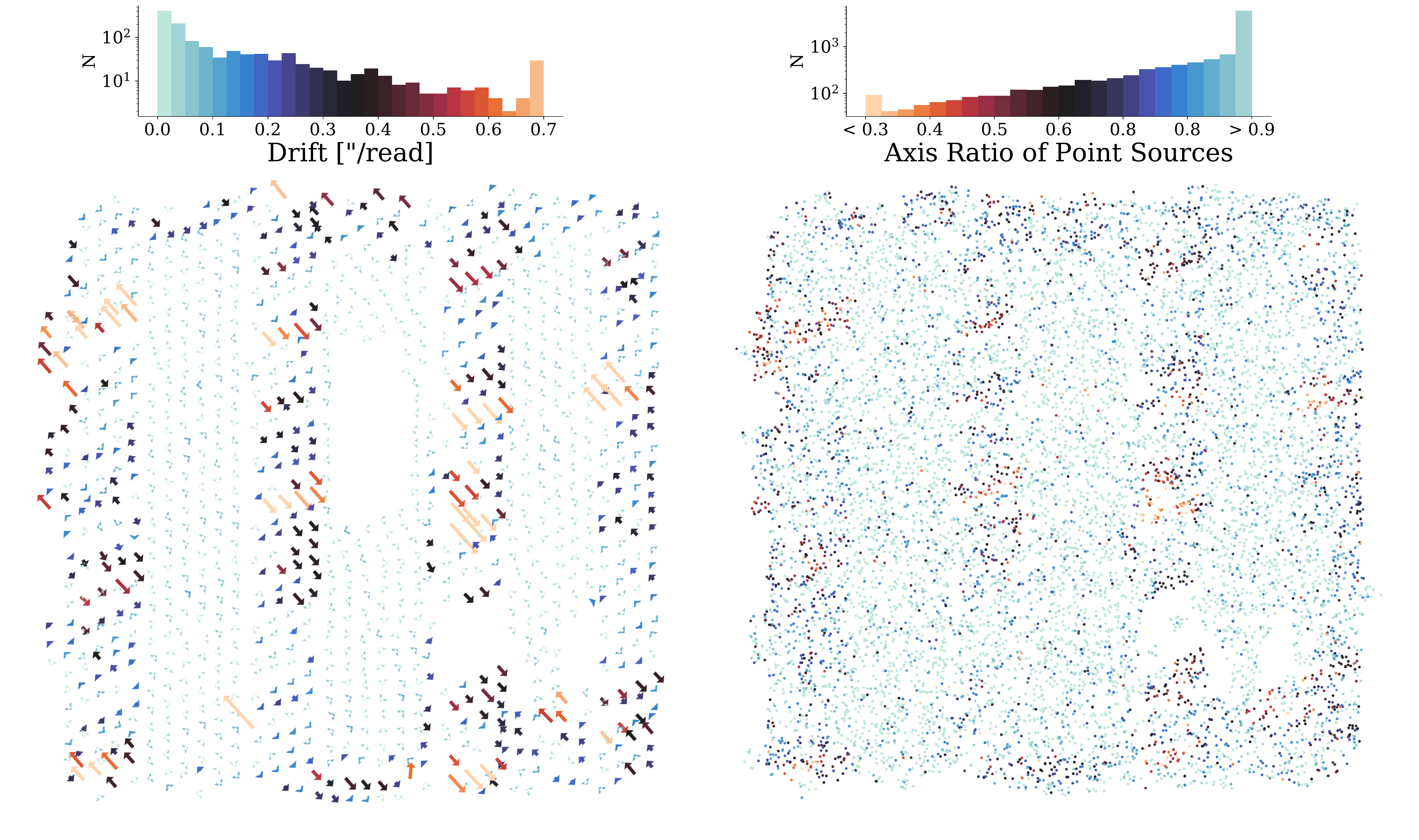}
\caption{\textit{Left:} The total drift of the telescope during an unguided exposure, with the length of the arrow representing the drift in one pointing and the color representing the average drift per read. The histogram on top shows the distribution of drifts in 3D-DASH. \textit{Right:} The axis ratio of point sources in the 3D-DASH image. The point sources were identified from the Hyper Suprime-Cam Subaru Strategic Program \citep{Aihara2021} and their axis ratios are measured using GALFIT. The point sources in the final image are elongated in regions of higher drift.} 
\label{fig:drift}
\end{figure*}

\subsection{Point Spread Function of 3D-DASH}
\label{sec:psf}


We created a grid of point spread functions (PSF) for the 3D-DASH mosaic using Grizli\footnote{\url{https://grizli.readthedocs.io}} to shift and drizzle HST empirical PSFs \citep{Anderson2015} at the positions of \npsfstar\ stars brighter than $I<25$ from the Hyper Suprime-Cam Subaru Strategic Program \citep{Aihara2021}. These \npsfstar\ stars sample the entire 3D-DASH mosaic area (see right panel of Figure \ref{fig:drift}), including the CANDELS region. We also provide a PSF generator tool\footnote{\url{www.lamiyamowla.com/3d-dash}} to determine the PSF at any location within the 3D-DASH footprint. 

Example position-dependent empirical PSFs and the stars at their positions are shown in Figure\ \ref{fig:psf}. While the position angle of each pointing is largely aligned, there are subtle variations visible in Figure 3 that motivate our decision to adopt a position-dependent PSF.   When comparing the orientation of star spikes across the mosaic in Figure 5b, we further see several cases where multiple orientation angles are co-added. The curves of growth, which show the fraction of light enclosed as a function of aperture size, for the DASH $H_{160}$ PSFs, normalized at 2$\arcsec$, are shown in the left panel of Figure \ref{fig:psf} for both the synthetic PSF and the empirical PSF from the stars. The PSFs are in agreement with each other and with the encircled energy as a function of aperture provided in the WFC3 handbook when also normalized to maximum radius of 2$\arcsec$. This demonstrates that the DASH technique does not introduce a significant smoothing.

To check the effect of the drift on the shape of the point spread functions, we measured the axis ratios in the \npsfstar\ stars using GALFIT \citep{Peng2010}. While most of the stars have axis ratios $>$ 0.8, we notice that stars in certain regions demonstrate flattening see \ref{fig:drift}. These regions coincide with regions of higher drift, which affected some orbits of Cycle 28. The regions of high drift ($> 0\farcs5$ per pointing) account for 5.4\% of the entire 3D-DASH area. This is reflected in the shape of the PSFs, with 4.3\% of PSFs having axis ratio $< 0.5$.

\subsection{Background Noise}
\label{sec:noise}

\begin{figure*}[ht]
\centering
\includegraphics[width=0.48\textwidth]{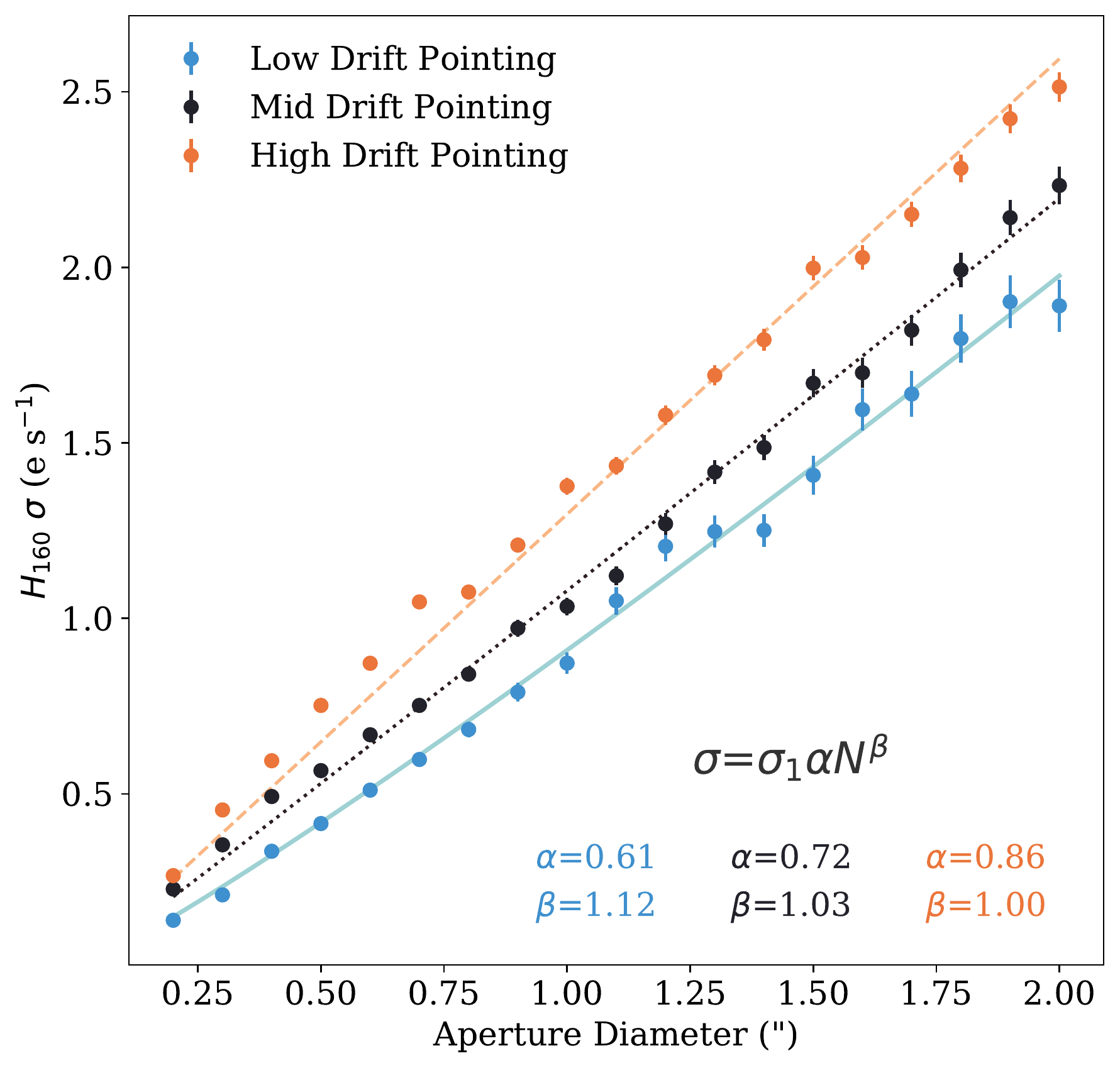}
\includegraphics[width=0.48\textwidth]{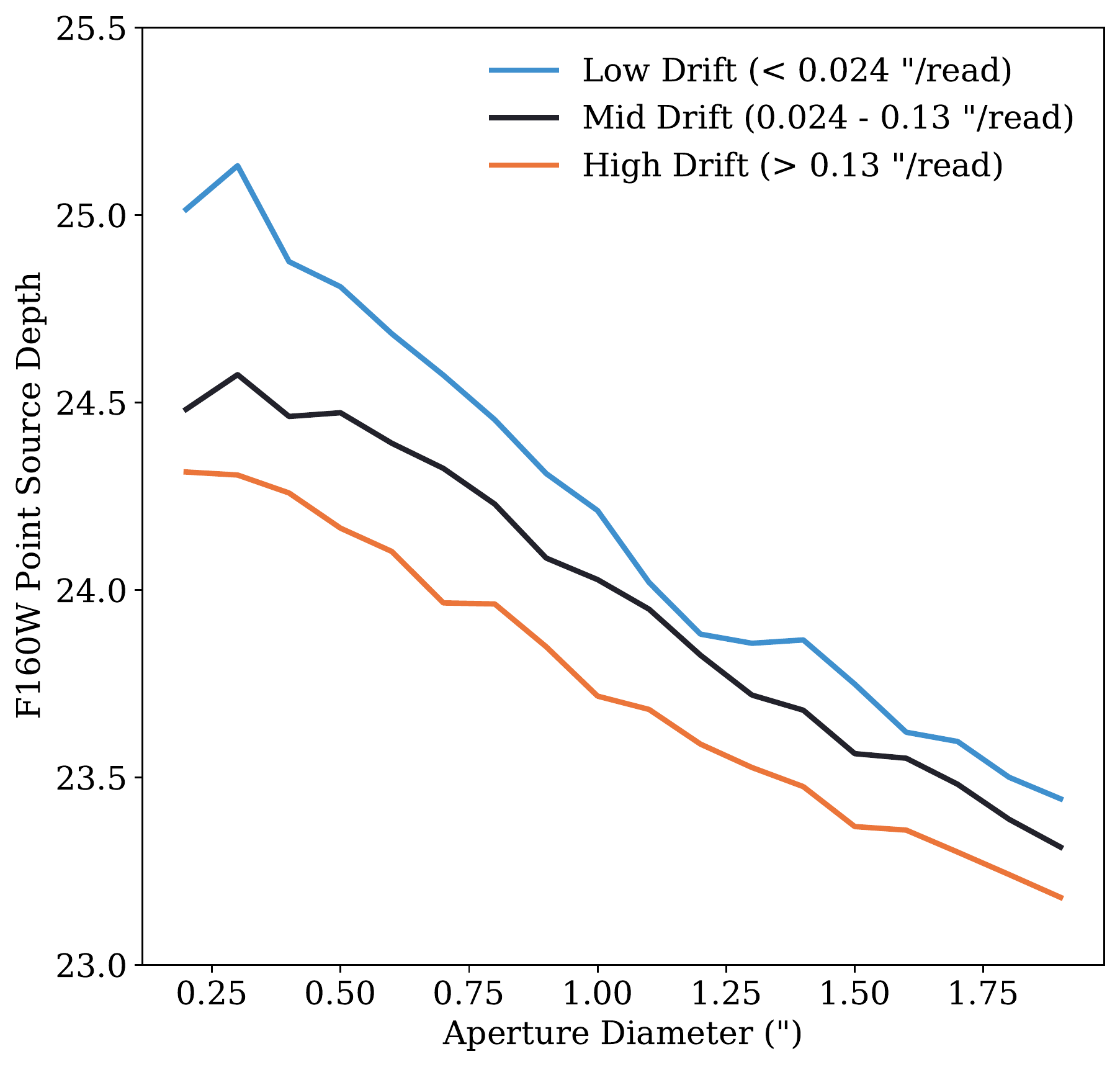}
\caption{The background noise level and the point source depth of 3D-DASH. To demonstrate the effect of the increased gyro drift in HST in Cycle 28, we split up our DASH-only pointings into three bins based on their drift rates: $dr<0\farcs024$ (low drift), $0\farcs24<dr<0\farcs13$ (medium drift), and $dr>0\farcs13$ per read (high drift). Left: Empty aperture photometry on the mosaic to determine the background noise level. The plot shows the standard deviation of the distribution of total flux in different aperture sizes measured from noise-equalized mosaic. The lines are power-law fits to the data, with the fit parameters shown below. The pointings with lower drifts demonstrate slightly steeper power-law  index that supports more correlated noise. Right: The point source depth measured in different aperture diameters. The low drift pointings reach depth similar to that of the pilot COSMOS-DASH, while high drift pointings result in shallower images. }
\label{fig:empty_ap}
\end{figure*}

The depth of the mosaic is determined by the background noise, which is expected to scale with the size of the aperture that is used for photometry. To determine how the background noise scales with aperture size, we measure the distribution of counts in empty regions of increasing size within the noise-equalized F160W image. For each aperture size we measure the flux in $10,000$ apertures placed at random positions across the combined 3D-DASH mosaic with both the DASH and the ``standard'' depth pointings. DASH depths are selected with weight$<100$ and ``standard'' depths have weight$>100$. Apertures that overlap with sources in the detection segmentation map are randomly moved until they no longer overlap with a source. Figure \ref{fig:empty_ap} shows the distribution of flux counts for increasing aperture size ranging from 0.2 to 2$\arcsec$ diameters in the 3D-DASH image. Each histogram can be well-described by a Gaussian, with the width increasing as aperture size increases. The increase in standard deviation with linear aperture size $N=\sqrt{A}$, where A is the area within the aperture, can be described as a power law. A power-law index of 1 would indicate that the noise is uncorrelated, whereas if the pixels within the aperture were perfectly correlated, the background noise would scale as $N^2$. The right-hand panel of  Figure \ref{fig:empty_ap} shows the measured standard deviation as a function of aperture size in the noise-equalized 3D-DASH image. We fit a power-law of the form
\begin{equation}
    \sigma = \sigma_1 \alpha N^{\beta},
\label{eq:empty_ap}
\end{equation}

where $\sigma_1$ is the standard deviation of the background pixels fixed to a value of 1.5 here, $\alpha$ is the normalization and $1<\beta<2$ \citep{Whitaker2012}. The fitted parameters are shown in Figure \ref{fig:empty_ap}. The power-law fit is shown by the solid line in the figure.

We also separate individual DASH apertures based on the program they were taken in (Cycle 23 COSMOS-DASH or Cycle 28 3D-DASH) to determine if the more recent DASH observations differ significantly in noise properties from the pilot program. On average, the 3D-DASH data with low drift rates similar to the pilot COSMOS-DASH data are deeper with a slightly steeper power-law  index that supports more correlated noise (see Figure \ref{fig:empty_ap}.  Those pointings with the highest drift instead have shallower depths (noisier data) and $\beta \sim1$, suggesting uncorrelated noise.  This makes sense as the drift rate in these cases is larger than the PSF itself between reads.


\subsection{Point Source Depth}
\label{sec:psd}

With the PSF and noise in hand we can calculate the photometric depth of the mosaic.  In the right panel of Figure\ \ref{fig:empty_ap} we show the $5\sigma$ depth as a function of aperture size. Apertures of a given size are placed at random positions within the mosaic, and we then calculate the $1\sigma$ variation in the measured fluxes within these apertures. A multiplicative aperture correction is applied based on the flux that falls outside that aperture for a point source, and multiplied by 5 to  estimate the 5$\sigma$ depth \citep[see][]{Skelton2014}. For very small apertures the S/N is suppressed because of the large aperture corrections that need to be applied, and for very large apertures the S/N is suppressed because of the high noise within the aperture \citep[see e.g., Figure 7 in][]{Whitaker2011}. The optimal aperture for HST NIR imaging is $0\farcs 3$, and the $5\sigma$ depth within that aperture is \psdepthdash\ for 3D-DASH pointings (depth at $0\farcs 7$ is 24.4$\pm$0.2). This can be compared to the expected depth for guided exposures of the same exposure time, as determined by the Exposure Time Calculator. The expected depth is
$\approx 25.2$, which means that use of the DASH technique imposes a penalty of $\approx 36$\,\% on the image depth due to increased gyro drift. 

The Cycle 28 3D-DASH has an average point-source depth $\sim0.4$ shallower than COSMOS-DASH. The cause of this increased noise and shallower imaging is likely increased drift rates in the DASH observations due to the new gyros. To check for this, we split the new DASH pointings into 3 equal sized bins based on drift rates: $dr<0\farcs024$ (low drift), $0\farcs24<dr<0\farcs13$ (medium drift), and $dr>0\farcs13$ per read (high drift) (see Figure \ref{fig:empty_ap}). The pointings with low drift rates reach a point source depth (H$_{160}\sim$25.1) similar to that of COSMOS-DASH, while the pointings with high drift rates produce shallower images (H$_{160}\sim$24.3). 

\begin{figure*}[ht]
\centering
\includegraphics[width=\textwidth]{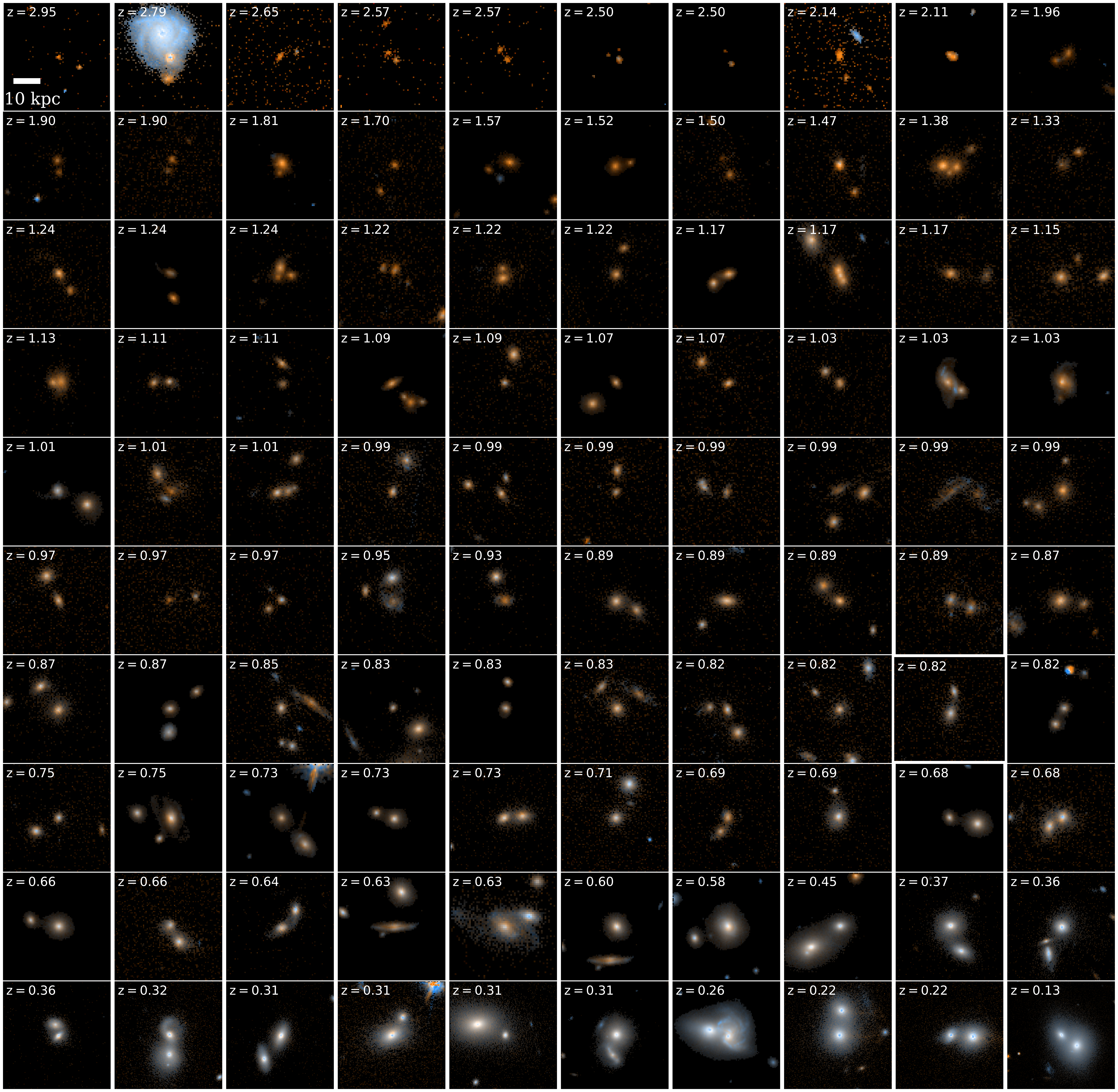}
\caption{With the wide-area high-resolution imaging we now have a census of rare close pair of galaxies, critical for studying evolution of galaxy merger rates. Stamps show close pair galaxies in the 3D-DASH field with $\log(M_\star/M_\odot) > 10.5$ at $z<3.0$ with $H<22$ and separation less than 20 kpc (2D separation with an additional cut of $z_{\rm diff}<0.05$) from the UVISTA catalog. The galaxies are sorted by their redshift, which are shown in the top left corner. Each stamp is 40 kpc $\times$ 40 kpc and are created using DASH-area only WFC3/F160W and ACS/F814W images.}
\label{fig:closepairs}
\end{figure*}

\begin{figure*}[ht]
\centering
\includegraphics[width=\textwidth]{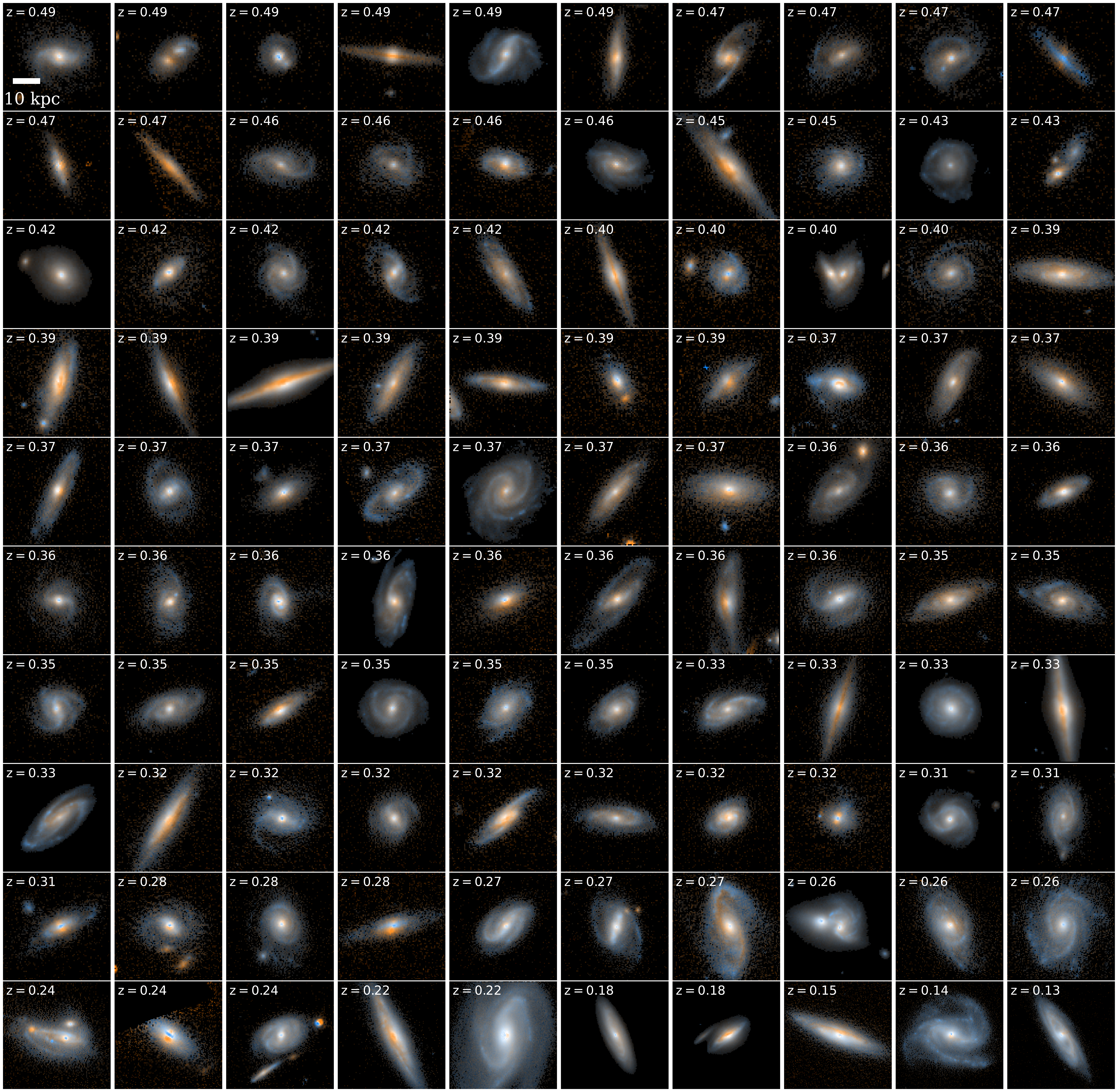}
\caption{The wide area imaging combined with grism spectroscopy, gives a census of the morphology of star-forming regions of the rare massive star-forming galaxies in the last 5 Gyrs. The stamps show the most massive star-forming galaxies (selected with $U-V$ and $V-J$ color cuts from \citet{Whitaker2011}) in the 3D-DASH field with $\log(M_\star/M_\odot) > 10.8$ at $z<0.5$ from the UVISTA catalog. The galaxies are sorted by their redshift, which are shown in the top left corner. Each stamp is 40 kpc $\times$ 40 kpc and are created using DASH-area only WFC3/F160W and ACS/F814W images.}
\label{fig:sfmassive}
\end{figure*}

\begin{figure*}[ht]
\centering
\includegraphics[width=\textwidth]{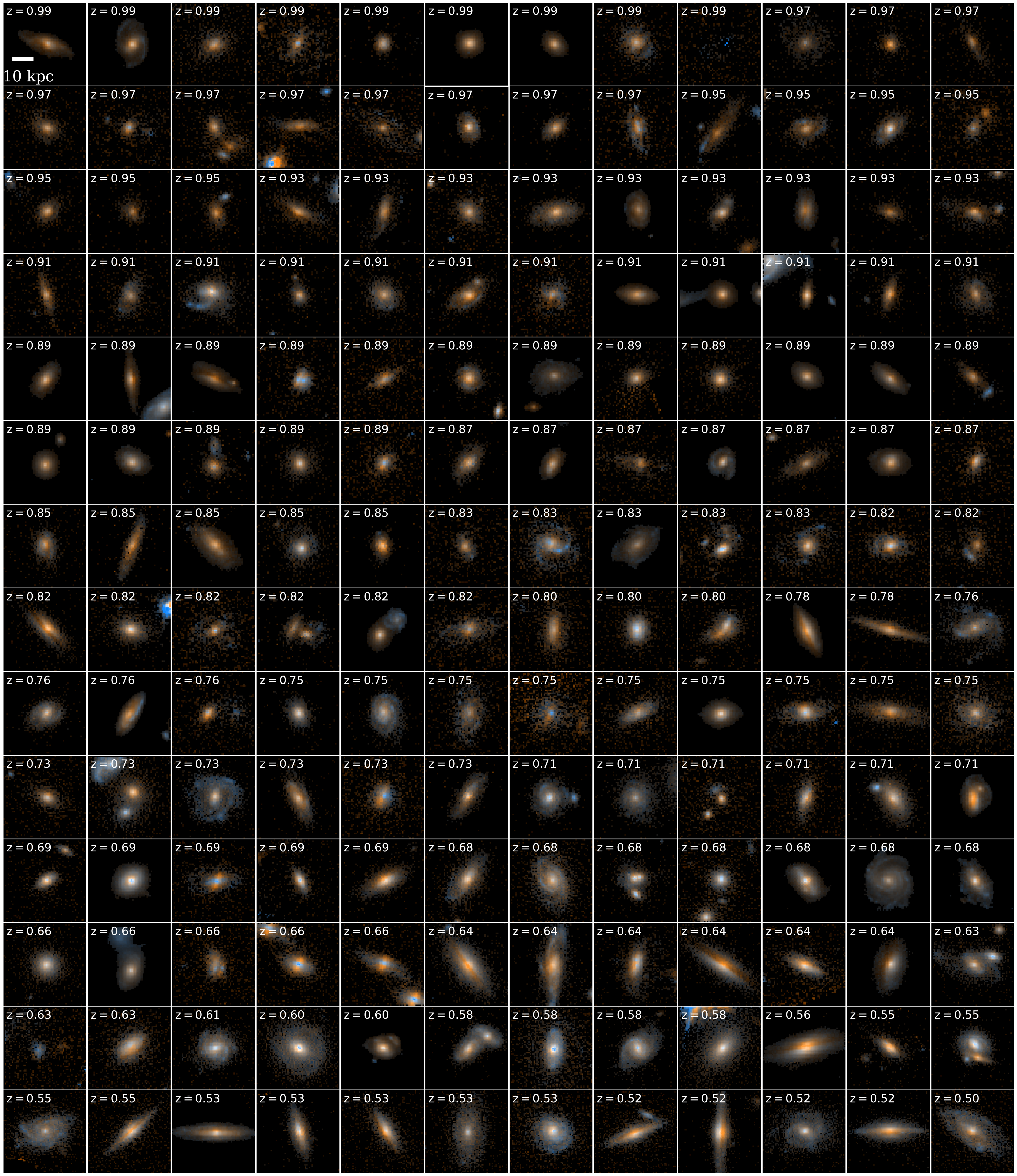}
\caption{The wide area imaging combined with HST grism spectroscopy, deep ground-based multiwavelength imaging \citep[UVISTA][]{Muzzin2013} and spectroscopy \citep[LEGA-C][]{vanderwel2016},  gives a census of the evolution of star-forming regions of the rare massive star-forming galaxies at $0.5<z<1$. The stamps show the most massive star-forming galaxies (selected with $U-V$ and $V-J$ color cuts from \citet{Whitaker2011}) in the 3D-DASH field with $\log(M_\star/M_\odot) > 11$ at $0.5<z<1.0$ from the UVISTA catalog. The galaxies are sorted by their redshift, which are shown in the top left corner. Each stamp is 40 kpc $\times$ 40 kpc and are created using DASH-area only WFC3/F160W and ACS/F814W images.}
\label{fig:legacmassive}
\end{figure*}

\section{Example Images and Science Cases}
\label{sec:example}

The purpose of the present paper is to describe the 3D-DASH program and its first major data product, a WFC3/IR $H_{160}$ mosaic of the COSMOS field. Besides the 3D-DASH data themselves we include data from 18 previous imaging programs in the mosaic, including the Cycle 23 COSMOS-DASH pilot program.
These data are not deep by conventional HST standards: the exposure time at most positions in the mosaic is only $\approx 250$\,s. Nevertheless, thanks to HST's superb resolution and the low background from space, the resulting $5\sigma$ point source depth of \psdepthdash\ (0\farcs3 diameter aperture) and 24.4$\pm$0.2 (0\farcs7 diameter aperture) adds a high resolution shallow tier to the deep observations from the ground (e.g. NEWFIRM Median Band Survey \citet{Whitaker2011}, UVISTA \citet{Muzzin2013}, Hyper Suprime-Cam Subaru Strategic Program \citet{Aihara2021}) .

The data have a wide range of science applications, and are particularly powerful for studies that are currently not limited by depth but by resolution in the NIR. In Figs.\ \ref{fig:closepairs}, \ref{fig:sfmassive} and \ref{fig:legacmassive}, we show example galaxy images for three such projects. 

\subsection{Massive Close Pairs}
In Figure\ \ref{fig:closepairs} we show massive close pairs, many of which are expected to be mergers in progress that lead to the build-up of today's giant elliptical galaxies \citep[see, e.g.,][]{delucia2007,Bezanson2009,Naab2009,Cooke2019}.
For this example, we show 3D-DASH images of galaxies selected from the ground-based UVISTA catalog of \citet{Muzzin2013} that satisfy the following conditions: $\log(M_\star/M_\odot) > 10.5$ at $z<3.0$ with $H<22$ and separation less than 20 kpc (2D separation with an additional cut of $z_{\rm diff}<0.05$). 

These galaxies are just-resolved in ground-based data, but it is not possible to measure the properties of the individual objects. In the 3D-DASH dataset they are clearly distinct, showing a variety of colors and morphologies. The minimum separation of close pairs in the UVISTA is set by the ground-based resolution, and \citet{Marsan2019} find that up to 1/3 of the most massive single ground-selected massive galaxies are in fact close pairs at HST resolution. 
This is mitigated by the 3D-DASH images, which provide resolution high enough to deblend the close pairs at low redshifts. This has significant implications for the high-mass end of low-z ground based mass functions, as well as the inferred evolution of galaxies from techniques such as abundance matching \citep[e.g.][]{Behroozi2013} or empirical methods \citep[][]{Moster2018, Behroozi2019}.
In addition, current measurements of the massive galaxy merger rate at $z>>1$ are still uncertain and model-dependent \citep[e.g.][]{Duncan2019,Pena2021}. A 3D-DASH-selected catalog will identify a complete sample of apparent massive galaxy close-pairs, further constraining the contribution of mergers to the build-up of massive galaxies.

\subsection{Massive Star-Forming Galaxies}

A second example is shown in Figure\ \ref{fig:sfmassive}. These are relatively bright and large galaxies at $z<0.5$ that are easily detected and studied from the ground. While these low redshift galaxies are spatially-resolved enough from the ground for simple morphological measurements such as size and compactness, they are not resolved enough for detailed morphological analysis pertaining to their growth. The combination of $H_{160}$ and $I_{814}$ from 3D-DASH and ACS-COSMOS provides spatially-resolved morphological information at two distinct wavelengths and enable the identification of star-forming knots and dust lanes. 

The increased spatial resolution also allows for the identification and analysis of clumpy star formation \citep[][]{Guo2018, Huertas2020}. The characteristics of clumps, including numbers and lifetimes, as well as morphological features like tidal tails have important implications for the strength and mechanisms of stellar feedback and galactic winds \citep[][]{Elmegreen2021, Dekel2022}.

\subsection{Cosmic Noon Counterparts of Massive Galaxies}

A third example is shown in Figure\ \ref{fig:legacmassive}, which are the intermediate redshift counterparts of star-forming massive galaxies shown in Figure\ \ref{fig:sfmassive}. These are relatively bright and large galaxies at $0.5<z<1.0$ that have deep spectroscopic observations from the ground in the LEGA-C survey \citep{vanderwel2016} but which are not spatially-resolved in the deep ground based NIR imaging \citep[UVISTA][]{Muzzin2013}. At $z>0.5$ the $I-H$ colors are critical for interpreting the morphologies, as they enable the identification of dust, star forming regions, and old stellar populations and -- more generally -- the conversion of light-weighted to mass-weighted images  \citep{Forster2011,Wuyts2012,Suess2019}.
These images can be combined with the myriad of spectroscopic programs in the COSMOS field \citep[such as][]{vanderwel2016}, for a comprehensive description of the build-up of the Hubble sequence over the past 5\,--\,8 Gyr. 

\section{Summary and Outlook}
\label{sec:summary}

In 2015 we demonstrated the capability of HST to observe wide area efficiently using the Drift And SHift technique. It was implemented in the COSMOS-DASH program which imaged \cdasharea\ deg$^2$ of the three deep stripes of the UltraVISTA fields. 3D-DASH extends this program by adding grism spectroscopic observations and by imaging the remaining area of the ACS-COSMOS field. Here, we combine the newly obtained 3D-DASH pointings with archival data in the COSMOS field to produce a \totarea\ deg$^2$ mosaic in the $H_{160}$ filter. 
This is the largest area ever imaged in the near-infrared at this spatial resolution, a situation that will likely not change until the launch of Euclid \citep{Euclid2011} and Roman \citep{wfirst12012,wfirst22015}. 

We show that the point source depth of the mosaic is \psdepthcomb\, shallower than pointed observations, and that the depth correlates with the size of the drifts between exposures.
As discussed in Section \ref{sec:psf}, the point spread function is only slightly degraded after shifting and adding the exposures. We provide two PSFs for each position: one that is noiseless and only includes the theoretical position-dependence, and a noisier empirical one that also includes the effects of in-sample drift and imperfect alignments. We recommend that researchers use both PSFs in their analyses and include any differences in outcomes in their error budgets. All our data products are available at MAST as a High Level Science Product via \dataset[10.17909/srcz-2b67]{\doi{10.17909/srcz-2b67}} and here \url{https://archive.stsci.edu/hlsp/3d-dash/}. We also provide tools to generate local point spread functions and make cutouts at any location within the 3D-DASH footprint\footnote{\url{www.lamiyamowla.com/3d-dash}}. 

We are publicly releasing this image which can be used as a complement to other datasets in the field, such as the LEGA-C spectroscopic survey \citep{vanderwel2016}, or
serve as the primary data for studies that require high resolution data in the NIR. We anticipate that these data will also be helpful for planning JWST observations of objects in the COSMOS field.
In the future we plan to provide an $H_{160}$-selected catalog of objects in the mosaic as well as deliver the grism observations that are part of the 3D-DASH project.

\begin{deluxetable*}{cllcc}
\label{tab:struct}
\tablecaption{Structure of a Single GO-16259 Orbit.}
\tabletypesize{\footnotesize}
\tablewidth{0pt}
\tablehead{\colhead{Step} & \colhead{Event} & \colhead{Changes to keywords} &
\colhead{Duration} & \colhead{Exposure time}}
\startdata
1 & guide star acquisition  & \nodata & 383\,s & \\
2 & guided exposure, position 1 & {\tt PCS Mode=FINE} & 295\,s & 253\,s \\
   & & {\tt SAMP-SEQ=SPARS50} & & \\
   & & {\tt NSAMP=6} & & \\
3 & stop FGS corrections & \nodata  & 21\,s &\\
4 & offset to position 2 & \nodata  & 59\,s & \\
5 & unguided exposure, position 2 & {\tt PCS Mode=GYRO} & 245\,s & 203\,s \\
   & & {\tt SAMP-SEQ=SPARS25} & & \\
   & & {\tt NSAMP=9} & & \\
6 & offset to position 3 & \nodata  & 59\,s & \\
7 & unguided exposure, position 3 & {\tt NSAMP=9}   & 245\,s & 203\,s \\
8 & offset to position 4 &  \nodata & 59\,s & \\
9 & unguided exposure, position 4 &   {\tt NSAMP=10} & 270\,s & 228\,s \\
10 & offset to position 5 &  \nodata & 58\,s & \\
11 & unguided exposure, position 5 & {\tt NSAMP=11} & 295\,s & 253\,s \\
12 & offset to position 6 &  \nodata & 59\,s & \\
13 & unguided exposure, position 6 &  {\tt NSAMP=12}  & 320\,s & 278\,s \\
14 & offset to position 7 &  \nodata & 59\,s & \\
15 & unguided exposure, position 7 &   {\tt NSAMP=8} & 220\,s & 178\,s \\
16 & offset to position 8 & \nodata  & 58\,s & \\
17 & unguided exposure, position 8 & {\tt NSAMP=8}   & 220\,s & 178\,s \\
\hline
& Unused orbital visibility:  & & 0\,s & \\
& Total duration and exposure time: &  & 2925\,s & 1774\,s 
\enddata
\end{deluxetable*}

\begin{table*}[]
	\centering
	\caption{Data included in the final H$_{160W}$ 3D-DASH$\_v1.0$ mosaic.  }
	\begin{tabular}{p{0.08\linewidth}  p{0.07\linewidth}  p{0.08\linewidth}  p{0.5\linewidth}  p{0.18\linewidth}}
		\hline
		\hline
		\textbf{Program Number} &\textbf{HST Cycle} & \textbf{Number of Pointings}& \textbf{Program Name}& \textbf{Principal Investigator}  \\ \hline
		\textbf{12167} & 18 & 28& Resolving the Matter of Massive Quiescent Galaxies at $z=1.5-2$ & Marijn Franx  \\
		\textbf{12440} & 19 & 176 & Cosmic Assembly NIR Deep Extragalactic Legacy Survey -- GOODS-South Field, Non-SNe-Searched Visits & Sandra Faber, Henry Ferguson\\
		\textbf{12461}& 19 & 23 & Supernova Follow-up for MCT & Adam Reiss  \\
	\textbf{12578} & 19 & 14 & Constraints on the Mass Assembly and Early Evolution of $z\sim2$ Galaxies: Witnessing the Growth of Bulges and Disks & Natascha F\"orster Schreiber \\
	\textbf{12990} & 20 & 52& Size Growth at the Top: WFC3 Imaging of Ultra-Massive Galaxies at $1.5<z<3$ & Adam Muzzin  \\
	\textbf{13294} & 21 & 6 & Characterizing the formation of the primordial red sequence & Alexander Karim  \\
	\textbf{13384} & 21  & 4& A Simultaneous Measurement of the Cold Gas, Star Formation Rate, and Stellar Mass Histories of the Universe & Dominik Riechers \\
	\textbf{13641} & 22  & 36& A Detailed Dynamical And Morphological Study Of $5<z<6$ Star, Dust, and Galaxy Formation With ALMA And HST & Peter Capak \\
\textbf{13657}& 22 & 116& Probing the Most Luminous Galaxies in the Universe at the Peak of Galaxy Assembly & Jeyhan Kartaltepe   \\
\textbf{13868} & 22 & 44 & Are Compton-Thick AGN the Missing Link Between Mergers and Black Hole Growth? & Dale Kocevski \\
\textbf{14114} & 23 & 456& A Wide-Field WFC3 Imaging Survey in the COSMOS Field & Pieter van Dokkum  \\
\textbf{14721} & 24 & 40& The Fundamental Plane of Ultra-Massive Galaxies at $z\sim2$ & Christopher Conselice  \\
\textbf{14750}& 24 & 8 & Exploring environmental effects on galaxy formation with WFC3 in the most distant cluster at $z=2.506$ & Tao Wang \\
\textbf{14895} & 24 & 20& Confirmation of ultra-luminous $z\sim9$ galaxies & Rychard Bouwens  \\ 
\textbf{15229} & 25 & 32& Spectroscopic redshifts and age dating of a first statistical sample of passive galaxies at $z\sim3$ & Emanuele Daddi   \\
\textbf{15692} & 26 & 24& HST imaging for an immediate study of the ISM in $z=4.5$ galaxies & Andreas Faisst   \\
\textbf{15910} & 27 & 25&  Galaxy evolution in a massive $z=2.91$ halo fed by cold accretion & Emanuele Daddi   \\
\textbf{16081} & 27 & 8 &Let's Point Hubble at a Bubble! & James Rhoads  \\
\textbf{16259}& 28 & 800 & 3D-DASH: A Wide Field WFC3/IR Survey of COSMOS & Ivelina Momcheva 
		\\\hline
		\hline
	\end{tabular}
	\label{tab:archival}
\end{table*}

\facilities{HST(STIS)}


\software{Astropy \citep{astropy:2013, astropy:2018},  \citep[\texttt{grizli}][]{grizli2019},  \citep[\texttt{Astrodrizzle}][]{Gonzaga2012},  \citep[\texttt{GALFIT}][]{Peng2010},  \citep[\texttt{Montage}][]{Montage2010}}





\begin{acknowledgments}
This paper is based on observations made with the NASA/ESA \textit{HST}, obtained at the Space Telescope Science Institute, which is operated by the Association of Universities for Research in Astronomy, Inc., under NASA contract NAS 5-26555. Support from NASA STScI grants HST-GO-14114, HST-GO-16259, and HST-GO-16443 is gratefully acknowledged. L.M. and K.I. gratefully acknowledges support from the Dunlap Institute for Astronomy and
Astrophysics through the Dunlap Postdoctoral Fellowship. The Dunlap Institute is funded through an endowment established by the David Dunlap family and the University of Toronto S.C. and K.W. wish to acknowledge funding from the Alfred P. Sloan Foundation and HST-AR-15027. K.W. acknowledge support under NASA Grant Number 80NSSC20K0416. The Cosmic Dawn Center is funded by the Danish National Research Foundation under grant No. 140.
\end{acknowledgments}

\bibliographystyle{aasjournal}
\bibliography{dash_bib.bib}



\end{document}